\documentclass[runningheads]{llncs}

\usepackage[T1]{fontenc}
\usepackage{graphicx}
\usepackage{listings}
\usepackage[scaled=0.85]{beramono}
\usepackage{enumitem}
\usepackage{booktabs}
\usepackage{tabularx}
\usepackage{xparse}
\usepackage{amsmath}
\usepackage{xspace}
\usepackage[skins]{tcolorbox}
\usepackage{cite}
\usepackage{hyperref}
\usepackage{url}
\DeclareUrlCommand\repo{\urlstyle{tt}}

\lstdefinestyle{promptblock}{
  basicstyle=\ttfamily\footnotesize,
  numbers=left,
  numberstyle=\tiny,
  stepnumber=1,
  numbersep=6pt,
  frame=single,
  breaklines=true,
  breakatwhitespace=true,
  columns=fullflexible,
  keepspaces=true,
  showstringspaces=false,
  xleftmargin=2em,
  framexleftmargin=2em,
  aboveskip=4pt,
  belowskip=4pt,
}
\lstdefinelanguage{Prompt}{
  moredelim=**[is][\itshape]{/*}{*/},
}
\newcommand{\bench}{\textsc{Vul4Py}\xspace}

\begin{document}

\title{\bench: Benchmarking Automated Vulnerability Repair in Python with Paired Exploit and Functional Oracles}
\titlerunning{Vul4Py}
\author{
  Tan Bui\inst{1} \and 
  Ting Zhang\inst{2}\thanks{Ting Zhang is the corresponding author.} \and 
  Ferdian Thung\inst{1} \and 
  Yunpeng Xiong\inst{2} \and 
  Penghao Jiang\inst{3} \and 
  Xin Zhou\inst{1} \and 
  David Lo\inst{1}
}

\authorrunning{T. Bui et al.} % Optional: Adjust short author list for running heads

\institute{
  Singapore Management University \\
  \email{\{ngoctanbui,ferdiant.2013,xinzhou.2020,davidlo\}@smu.edu.sg} 
  \and
  Monash University \\
  \email{\{ting.zhang,nemo.xiong\}@monash.edu}
  \and
  University of New South Wales \\
  \email{\{penghao.jiang\}@unsw.edu.au}
}

\maketitle

\begin{abstract}
% Software vulnerabilities are a critical software security challenge, and Automated Vulnerability Repair (AVR) has advanced across program analysis, machine learning, and Large Language Models (LLMs). 
% Python is among the most widely used programming languages and powers critical web, data, and ML infrastructure, so Python vulnerabilities are common and consequential in practice. 
% % We present \bench, the largest Python-only AVR benchmark that requires paired functional and exploit oracles on every entry: 100 real vulnerabilities from 60 open-source projects, spanning 60 distinct CWEs and the years 2017 to 2025, each shipping vulnerable and fixed revisions with a project-native pytest oracle that must fail before the fix and pass after it. 
% \bench contains 100 real vulnerabilities from 60 open-source projects, spanning 60 distinct CWEs and the years 2017 to 2025, each packaged with a pinned, reproducible environment.
% Using \bench, we compare six approaches across three categories: a specialized vulnerability repair tool, direct prompting LLMs, and software engineering agents. 
% Our experimental results show that the agents clearly dominate, with OpenHands the strongest, while direct prompting and the specialized tool repair far fewer vulnerabilities despite sharing the same backbone model. 
% Manual review further confirms that the paired oracle is a high-precision proxy for genuine repair.
% \bench is publicly available at \url{https://github.com/tabudz/vul4py}.

Automated Vulnerability Repair (AVR) has advanced rapidly across program analysis, machine learning, and Large Language Models (LLMs), but a verifiable, head-to-head comparison of AVR approaches on Python is still missing.
Python underpins critical web, data, and machine-learning infrastructure, yet existing Python benchmarks accept a patch on the strength of a proof-of-concept exploit alone, or apply a functional test only on the subset of entries whose upstream project happens to ship one. 
Both therefore miss \emph{functional regressions}, in which a patch defeats the exploit but breaks unrelated behavior.
We present \bench, a Python AVR benchmark in which every entry carries a \emph{paired oracle}: an exploit oracle that must fail on the vulnerable revision and pass on the fixed one, together with a project-native pytest functional oracle that must pass on both.
\bench comprises 100 real vulnerabilities from 60 open-source projects, spanning 60 distinct CWEs and the years 2017 to 2025, each packaged with a pinned, reproducible per-instance environment.
Using \bench, we compare six approaches in three categories: a specialized vulnerability repair tool, directly prompted LLMs, and software engineering agents. 
The agents dominate: OpenHands repairs 41 of 100 vulnerabilities, against 4 for the strongest directly prompted LLM and 2 for the specialized tool, despite all three sharing the same backbone model. 
The paired oracle is what makes these counts trustworthy: it rejects 15 of the 119 patches that an exploit-only oracle would accept, and 98 of the 104 patches it admits are manually confirmed to be semantically equivalent to the developer's patches.

\end{abstract}

\section{Introduction}
\label{sec:intro}

Software vulnerabilities are a critical software security concern, and a large body of work has gone into detecting them automatically~\cite{weyssow2025r2vul,zhou2024large,li2025out}. 
Detection, however, only localizes the exposure; removing it requires a patch, and once a vulnerability is reported, a timely repair is necessary to mitigate the damage. 
Automated Vulnerability Repair (AVR) has therefore emerged as a vital field~\cite{li2025sok,hu2025sok}. 
A typical AVR pipeline has three phases~\cite{li2025sok}: vulnerability analysis, patch generation, and patch validation. 
Most prior work concentrates on patch generation, treating AVR as a sequence-to-sequence task in which a new approach proposes the fixed version of a vulnerable program~\cite{nong2025appatch,zhou2025large,yang2026securepair}. 
Patch validation has drawn far less attention, yet it is what separates a patch that looks right from one that is demonstrably a repair, and it determines what a benchmark can measure at all.

Large Language Models (LLMs) and agents have since advanced many software engineering
tasks~\cite{zhang2025revisiting,xiong2026sifting}, and AVR approaches have begun to apply them.
For instance, APPATCH~\cite{nong2025appatch} combines vulnerability semantics reasoning with adaptive prompting to patch code without test inputs or model fine-tuning. 
More recently, general-purpose software engineering agents~\cite{wang2024openhands,gao2025trae,yang2024swe} have made a different capability available: rather than emitting a patch in one shot, they can execute
the project, observe failing tests, and revise their edits. 
This shift raises two questions at once. 
The first is how much the ability to iterate against execution feedback actually buys on real vulnerabilities.
The second concerns validation: an evaluation that checks only whether the exploit is defeated will silently credit patches that overfit the exploit.

We study both questions on Python. 
Python is among the most widely used programming languages in modern software
development~\cite{tiobe2025,stackoverflow2025,githuboctoverse2024}, and its
package ecosystem is deeply interconnected, so a flaw in a widely depended-on
package remains exploitable across its downstream consumers long after
disclosure~\cite{alfadel2023empirical}. Its weakness profile also differs from
the memory-safety-dominated profile of C and C++ that most AVR work targets,
running instead to unsafe deserialization, resource exhaustion, and sandbox
escape (Section~\ref{sec:construction}). 
Despite this, tooling for evaluating automated repair of Python vulnerabilities remains limited.

\textit{Limitations in existing research.} Two recent surveys of AVR appeared in top-tier security
venues~\cite{hu2025sok,li2025sok}, but neither closes the gap for Python: Li et
al.~\cite{li2025sok} evaluate LLM-based approaches only at the patch generation
step rather than over the entire workflow, while Hu et al.~\cite{hu2025sok}
compare specialized repair approaches, 7 on C and C++ and 2 on Java, but omit
LLM-based ones. Three recent benchmarks target executable validation directly,
and Table~\ref{tab:benchmarks} contrasts them with \bench.
VulnRepairEval~\cite{vulnrepaireval} validates 23 Python instances with a
dual-container differential between the vulnerable and fixed revisions, but uses
a proof-of-concept exploit alone and ships no functional test.
PatchEval~\cite{patcheval} provides 1{,}000 multi-language entries with about 230
executable in a Docker sandbox, but applies its functional test only when the
upstream project happens to supply one. 
CVE-Bench~\cite{wang2025cve} provides 509
multilingual entries with 124 executable Python ones, but is partially
open-sourced at the time of writing. The consequence is shared: a patch that
defeats the exploit while breaking unrelated functionality is judged successful.
These functional regressions are exactly the failure mode that agentic repair is
most likely to produce and that current Python benchmarks are least able to
catch.

\begin{table}[t]
\centering
\caption{Executable Python vulnerability repair benchmarks. \emph{Exploit} is a
security test that distinguishes the vulnerable from the fixed revision;
\emph{Functional} is a non-security test suite required to keep passing on both
revisions; \emph{Public} indicates whether the instances and their execution
environments are openly released. \bench is the only benchmark that requires
both oracles on every entry.}
\label{tab:benchmarks}
\small
\begin{tabular}{lrrccc}
\toprule
\textbf{Benchmark} & \textbf{Entries} & \textbf{Executable} & \textbf{Exploit} & \textbf{Functional} & \textbf{Public} \\
 & & \textbf{Python} & & & \\
\midrule
VulnRepairEval~\cite{vulnrepaireval} & 23      & 23                       & all & none          & yes \\
PatchEval~\cite{patcheval}           & 1{,}000 & n/r$^{\dagger}$          & all & conditional   & yes \\
CVE-Bench~\cite{wang2025cve}         & 509     & 124                      & all & not reported  & partial \\
\bench (this work)                   & 100     & 100                      & all & \textbf{all}  & yes \\
\bottomrule
\end{tabular}

\smallskip
{\footnotesize $^{\dagger}$ PatchEval reports about 230 executable entries across
all languages; the Python subset is not broken out.}
\end{table}
To close this gap, we introduce \bench, a Python AVR benchmark in which every entry carries a paired oracle.
Each of its 100 instances, curated from 60 open-source projects, ships a vulnerable and a fixed revision together with two executable pytest oracles: an \emph{exploit oracle}, derived from the security test that the upstream fix commit adds, which must fail on the vulnerable revision and pass on the fixed one; and a \emph{functional oracle}, the project's own non-security test suite, which must pass on both. 
An instance is admitted only if this before-and-after behavior reproduces, so a candidate patch is accepted only when it defeats the exploit \emph{and} leaves the rest of the project intact. 
To show how \bench is used, we evaluate six approaches in three categories: the specialized vulnerability repair tool
APPATCH~\cite{nong2025appatch}, two directly prompted LLMs, GPT-4o and Claude Sonnet 4, and three software engineering agents, OpenHands~\cite{wang2024openhands}, Trae Agent~\cite{gao2025trae}, and SWE-agent~\cite{yang2024swe}.

\textit{Contributions.} We summarize the contributions as follows:

\begin{itemize}[left=0pt, labelsep=0.5em, itemsep=0pt]
\item \textbf{Dataset.} We present \bench, the first Python AVR benchmark in which \emph{every} entry carries both an exploit oracle and a functional oracle. 
% It comprises 100 real vulnerabilities from 60 open-source projects, covers 60 distinct CWEs, and spans 2017 to 2025. \bench is built by a documented end-to-end curation pipeline (OSV ingestion, repository discovery, fix commit identification, automated \texttt{meta.json} generation, and manual oracle validation) and ships with an evaluation harness that parses pytest's JUnit XML output into structured per-test verdicts, so that genuine repair failures are automatically distinguished from infrastructure and dependency errors.

\item \textbf{Empirical study.} We conduct the first study on \bench comparing three repair paradigms under a matched interaction budget: a specialized vulnerability repair approach, directly prompted LLMs, and general-purpose software engineering agents. 
% The study establishes baselines for future Python AVR work on \bench.

% \item \textbf{Insights.} Software engineering agents outperform both direct prompting and the specialized tool by an order of magnitude despite sharing the same backbone model. 
\item \textbf{Insights.} Software engineering agents repair 22 to 41 of the 100 vulnerabilities, whereas the strongest non-agent approach repairs 4, despite all agents and the strongest baseline sharing the same backbone model.
Since the specialized tool relies on adaptive prompting and produces patches without interacting with the codebase, this gap suggests that the ability to execute code, run the test suite, observe failures, and iteratively edit the repository matters more for vulnerability repair than specialized prompting alone. 
We further show that the functional half of the oracle is not redundant: it rejects one in eight patches that an exploit-only oracle would have accepted.
\end{itemize}

\section{Background and Related Work}
\label{sec:background}

\subsection{Automated Vulnerability Repair}
AVR is a specialized branch of program repair focused on security flaws~\cite{hu2025sok}.
Existing approaches can be grouped as follows:
\begin{itemize}[left=0pt, labelsep=0.5em, itemsep=0pt]
\item \textit{Search-based and template-based methods}~\cite{le2011genprog,gao2019crash} modify vulnerable code by searching within a space of candidate patches or applying predefined fix templates such as adding input validation.
These methods can be effective for certain known vulnerability classes but may lack flexibility for novel problems~\cite{hu2025sok}.
\item \textit{Semantics-based methods} use program analysis, either static or dynamic, to guide patch generation. These approaches~\cite{shariffdeen2025vulnerability,zhang2022program,gao2021beyond} use techniques such as concolic execution to pinpoint the root cause of security failures and guide targeted code modifications.
Such methods typically identify failing execution paths that violate security assertions and search for patches that satisfy the necessary constraints, often by adding checks, correcting calculations, or restricting inputs to safe ranges.
\item \textit{Learning-based methods} typically use deep learning or LLMs to transform vulnerable code into non-vulnerable code.
Earlier work fine-tunes transformer-based models such as T5~\cite{raffel2020exploring} or CodeBERT~\cite{feng2020codebert} on vulnerability fix pairs.
These models~\cite{fu2022vulrepair,zhou2024out,chen2022neural} can capture common fix patterns but depend on the availability of large curated vulnerability datasets.
They also tend to be language-specific and struggle if the vulnerability type was not seen in training.
\end{itemize}

\subsection{Repair via Large Language Models and Agents}
\textbf{LLMs} have emerged as powerful resources for various software engineering tasks such as code completion~\cite{husein2025large}, code summarization~\cite{sun2024source}, and program repair~\cite{zhang2024systematic}.
For vulnerability repair, the key question is simple: given only the vulnerable snippet, can an LLM spot the issue and suggest a correct fix?
Unlike traditional vulnerability repair approaches, an LLM does not need a test case or a formal specification.
Instead, it relies on what has been learned about code semantics and common vulnerability patterns~\cite{pearce2023examining}.
This is especially useful when details such as the exact CWE or exploit are unknown, which is common in practice~\cite{mell2024measuring}.
Many existing vulnerability repair approaches also assume prior knowledge of the exact location or type of the vulnerability~\cite{shariffdeen2025vulnerability,fu2022vulrepair,chen2022neural}.
In contrast, LLMs can often work directly from source code input, making them more flexible.

However, harnessing LLMs for vulnerability repair requires careful prompt design.
Naively asking a model to \texttt{fix the vulnerabilities in this code} may yield an incomplete or incorrect patch, because code generation by LLMs is sensitive to how requirements are phrased~\cite{paleyes2025prompt}.
Effective prompts for vulnerability repair using LLMs typically include a brief instruction such as \texttt{You are a precise software engineer and security patcher} and the entire vulnerable function with its dependencies or source code file.
In our study, we use a structured prompt template to ensure consistency across models.

\vspace{3px}
\noindent\textbf{LLM agents} represent a new frontier in automated repair.
An agent uses an LLM as the core reasoner and augments it with tool use via protocols such as Anthropic's Model Context Protocol~\cite{anthropic_mcp_2024_news}, which standardizes how applications expose tools, data sources, and prompts to the model.
Instead of a single prompt response, the agent runs in an iterative loop that interleaves reasoning and actions~\cite{yao2022react} and can execute supported tools~\cite{schick2023toolformer} or perform self-reflection over intermediate feedback~\cite{shinn2023reflexion}.
For software engineering, such agents can perform project-level debugging by navigating a codebase, localizing the fault, applying edits, and validating the result using the project's own execution signals~\cite{yang2024swe}.

% \paragraph{General-purpose software engineering agents.}
% In contrast to specialized vulnerability repair pipelines, 
General-purpose agents are typically not tied to a specific task and rely on project feedback such as compiler errors, failing tests, or runtime crashes, rather than vulnerability-specific instrumentation.
They are promising for vulnerability repair because they can operate at repository scope, iteratively narrowing down issues and testing candidate fixes end-to-end.
In this work, we evaluate three general-purpose software engineering agents, OpenHands~\cite{wang2024openhands}, Trae Agent~\cite{gao2025trae}, and SWE-agent~\cite{yang2024swe}, each of which provides autonomous code editing and command execution in a bounded interaction budget.

\subsection{Python Vulnerability Datasets and Benchmarks}
Several Python vulnerability datasets are widely used.
CVEfixes~\cite{bhandari2021cvefixes}, BigVul~\cite{fan2020ac}, PatchDB~\cite{patchdb}, and VulnPatchPairs~\cite{vulnpatchpairs} are diff-paired corpora: each entry is a vulnerable--fixed snippet pair, suitable for similarity-based evaluation but not for executing a candidate patch against tests.
Project-level Python debugging benchmarks such as BugsInPy~\cite{widyasari2020bugsinpy} target functional bugs rather than vulnerabilities and lack exploit oracles.

% Three benchmarks are closer to \bench in spirit.
% VulnRepairEval~\cite{vulnrepaireval} provides 23 Python instances and validates with a dual-container differential between the vulnerable and fixed revisions, but uses only a proof-of-concept exploit, with no functional test.
% PatchEval~\cite{patcheval} provides 1{,}000 multi-language entries with about 230 in a Docker sandbox. Its oracle is the pair of \texttt{fix\_run.sh} and \texttt{vul\_run.sh} scripts for the exploit, and a conditional \texttt{unit\_test.sh} that is applied only when present in the project.
% CVE-Bench~\cite{wang2025cve} provides 509 multilingual entries with 124 executable Python ones, but is partially open-sourced.
% The distinguishing property of \bench is the paired functional and exploit oracle on every entry. PatchEval applies the functional test only when an upstream \texttt{unit\_test.sh} exists, while VulnRepairEval has no functional test at all. 

Three benchmarks are closer to \bench in spirit, summarized in Table~\ref{tab:benchmarks};
what separates them is how the functional half of the oracle is treated.
VulnRepairEval~\cite{vulnrepaireval} validates a candidate patch with a dual-container
differential between the vulnerable and fixed revisions, but the differential is driven by a
proof-of-concept exploit alone and no functional test is run at all.
PatchEval~\cite{patcheval} pairs \texttt{fix\_run.sh} and \texttt{vul\_run.sh} for the exploit
with a \texttt{unit\_test.sh} that executes only when the upstream project happens to ship one,
so the functional check is conditional on the subject project rather than guaranteed by the
benchmark. CVE-Bench~\cite{wang2025cve} does not report a functional criterion and is only
partially open-sourced. In \bench, both oracles are an admission precondition, so every entry
carries them.

\section{Constructing \bench}
\label{sec:construction}

\begin{figure}[t]
\centering
\includegraphics[width=\linewidth]{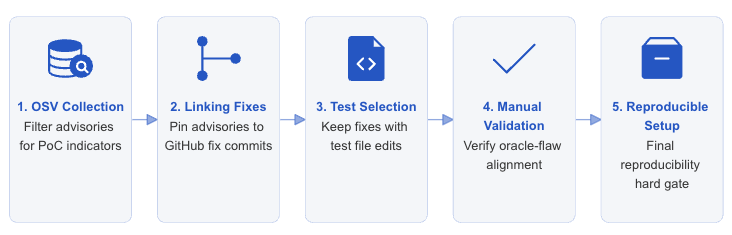}
\caption{The \bench construction pipeline. Open-source advisories from OSV are linked to fix commits, filtered to fixes that add or modify tests, manually validated through Manual Oracle Validation, and packaged into reproducible per-instance workspaces, each carrying a paired functional and exploit oracle.}
\label{fig:pipeline}
\end{figure}

We construct \bench from real Python vulnerability reports with executable pytest-based validation oracles.
Figure~\ref{fig:pipeline} shows how we build \bench using the pipeline.

\subsection{Candidate Collection from OSV}
We start from the Open Source Vulnerabilities database~\cite{osv} and retain Python advisories that
carry a proof-of-concept (PoC) indicator, such as a PoC keyword or a public
exploit reference. An executable benchmark needs a concrete trigger for every
entry, so advisories with no evidence of a reproducer are unlikely to yield an
exploit oracle and are filtered out before any further curation.

\subsection{Linking to Fix Commits}
% For each retained advisory, we follow its references to a GitHub fix commit,
% discarding cases that cannot be pinned to exactly one. 
For each retained advisory, we check its references for a link to the GitHub commit that fixed the vulnerability. 
We keep an advisory only if it can be traced to exactly one fix commit, and discard it otherwise.
A precise fix commit is
what lets us produce the paired vulnerable and fixed revisions on which the
benchmark rests; without an exact code transition there is no before-and-after
state against which to judge a candidate patch.

\subsection{Selecting Fixes with Tests}
Among the linked fixes, we keep those that add or modify test files.
This is the design choice at the heart of \bench: a test shipped by the fix commit is a developer-authored, vulnerability-revealing oracle, so prioritizing fixes that add or modify tests lets every entry carry an oracle grounded in the project's own conventions rather than one we create.

\subsection{Manual Oracle Validation}
A fix that touches a test file does not guarantee that the test detects the vulnerability. 
We therefore manually confirm a clear correspondence between each added or modified test and the reported flaw, discarding unrelated refactors, renamed fixtures, and coincidental test edits. 
Rather than a simple accept or reject, each candidate is judged against three explicit criteria:

\begin{itemize}[left=0pt, labelsep=0.5em, itemsep=0pt]
\item \textbf{Intent of test assertions.} The assertions must check the security-relevant behavior, for example that malicious input is rejected, sanitized, or raised, rather than merely confirming that a new feature works.
\item \textbf{Relationship to the code fix.} The test must exercise the code paths the fix changes, so that it fails on the vulnerable revision and passes on the fixed one for reasons attributable to the fix, not to incidental setup.
\item \textbf{Alignment with the OSV description.} The scenario the test constructs must match the vulnerability described in the advisory, so the oracle reflects the reported flaw.
\end{itemize}

% A candidate is accepted only if it satisfies all three, since an
% oracle that does not actually exercise the vulnerability would silently inflate
% the success of every approach.
A candidate is accepted only if it satisfies all three criteria. 
An oracle that does not test the actual vulnerability would let incorrect patches pass and make every approach appear more successful than it is.

\subsection{Reproducible Setup and Execution}
Finally, each accepted instance is packaged with an isolated per-instance
Python environment with pinned dependency
versions, together with the exact installation and test commands recorded in
its \texttt{meta.json}. An instance is admitted only if its oracle is
reproducible: the exploit test must fail on the vulnerable revision and pass on
the fixed revision, while the functional test passes on both. 

\subsection{Benchmark Characteristics}
\begin{figure}[t]
\centering
\includegraphics[width=\linewidth]{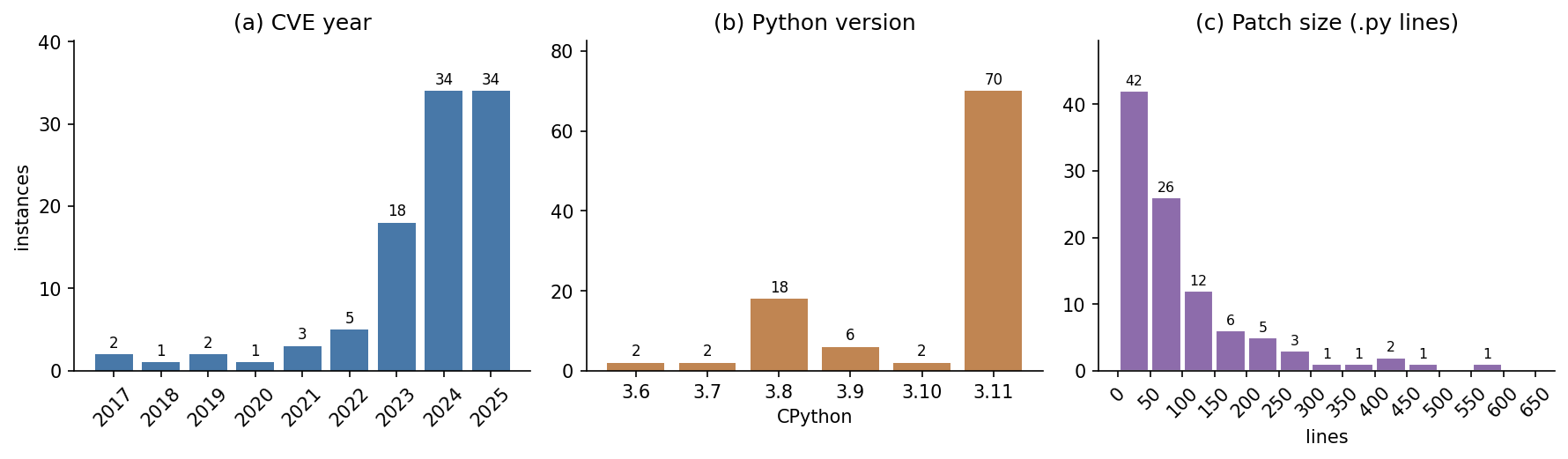}
\caption{Overall characteristics of the 100 \bench instances: (a) distribution by CVE year, (b) the Python version provisioned for each instance, and (c) patch size measured as \texttt{.py} lines changed by the upstream fix.}
\label{fig:dataset}
\end{figure}

We characterize the 100 \bench instances along five axes: temporal coverage, project diversity, vulnerability types, runtime environment, and patch size, with the overall picture summarized in Figure~\ref{fig:dataset}.

\textit{Year coverage.}
Instances span 2017 to 2025 with a heavy tail toward recent years: 34 instances are from 2025, 34 from 2024, and 18 from 2023, accounting for 86\% of the benchmark. The remaining 14 are distributed across 2017 to 2022. 

\textit{Projects.}
The 100 instances span 60 distinct open-source projects. The most represented projects are \repo{mmaitre314/picklescan} with 7 instances, \repo{pallets/jinja} with 5, and \repo{zopefoundation/RestrictedPython} and \repo{tornadoweb/tornado} with 4 each, followed by a cluster at 3 including \repo{mozilla/bleach}, \repo{urllib3/urllib3}, \repo{andialbrecht/sqlparse}, \repo{gitpython-developers/GitPython}, and \repo{latchset/jwcrypto}. The long tail of 39 projects with exactly one CVE is broad enough to discourage approach over-fitting.

\textit{CWE distribution.}
\bench covers 60 distinct CWE categories. Some instances are assigned more than one CWE, so the per-CWE counts sum to more than the 100 instances. The top categories by count are CWE-770, Allocation of Resources Without Limits, CWE-400, Uncontrolled Resource Consumption, and CWE-20, Improper Input Validation, each with 8 instances; CWE-22, Path Traversal, and CWE-79, Cross-Site Scripting, with 6; CWE-502, Deserialization, and CWE-345, Insufficient Verification of Data Authenticity, with 5. The distribution has a long tail: 42 CWEs are represented by exactly one instance.

\textit{Python version distribution.}
The Python versions provisioned for each instance span CPython 3.6 to 3.11: 3.11 dominates with 70 instances, followed by 3.8 with 18, 3.9 with 6, and 3.6, 3.7, and 3.10 with 2 each. Older Python interpreters are kept where the project's pinned dependencies do not install on 3.11.

\textit{Patch size.}
We measure patch size as total lines changed, that is, insertions plus deletions, by the upstream fix commit, restricted to non-test .py files.
The distribution is heavily right-skewed: median 65 lines, mean 103 lines, 10th to 90th percentile from 19 to 234 lines, and maximum 584 lines. 
Small fixes dominate; the few large fixes correspond to security-driven refactors such as complete rewrites of input parsers, while one fix changes no .py file at all because the vulnerability is resolved entirely outside Python source.

% Overall, \bench is recent without being narrow, broad in projects and weakness types without focusing in any one of them, and varied in fix complexity from single-line edits to multi-hundred-line refactors, all under reproducible, pinned runtime dependencies.

% \begin{tcolorbox}[enhanced, size=fbox, drop fuzzy shadow southeast, boxrule=0.1pt]
% Overall, \bench is a diverse, recent, and reproducible dataset: 100 instances over 60 projects and 60 CWEs spanning 2017 to 2025, with patch sizes from 0 lines to several hundred and pinned per-instance Python environments, which together make it a representative benchmark for Python AVR.
% \end{tcolorbox}

\section{Experimental Setup}
\label{sec:setup}

\subsection{Research Questions}
We aim to answer two research questions (RQs):

\begin{itemize}[left=0pt, labelsep=0.5em, itemsep=0pt]
\item \textbf{RQ1: How effective are current LLM-based approaches at repairing real Python vulnerabilities?} We compare APPATCH, GPT-4o, Claude Sonnet 4, OpenHands, Trae Agent, and SWE-agent under a common step limit, with a cost cap where the framework exposes one.

\item \textbf{RQ2: How reliable is the paired oracle as a proxy for correct
vulnerability repair?} We measure how many patches the functional oracle filters
out from those that pass the exploit test alone, and how precisely the paired
oracle predicts manually confirmed correctness.
\end{itemize}

\subsection{Evaluated Approaches}
We evaluate six approaches in three categories: a specialized vulnerability repair tool APPATCH~\cite{nong2025appatch}, two directly prompted LLMs, GPT-4o and Claude Sonnet 4, and three software engineering agents, i.e., OpenHands~\cite{wang2024openhands}, Trae Agent~\cite{gao2025trae}, and SWE-agent~\cite{yang2024swe}.

\subsubsection{Specialized AVR Tool.}
APPATCH is a recent patch generation system, assisted by an LLM, that combines adaptive prompting with vulnerability semantics reasoning. 
It extracts a vulnerability-focused program context via slicing and uses that context to guide an LLM to propose candidate fixes without any test cases or fine-tuning.

\textit{Implementation.} To evaluate APPATCH on Python, we adapted the original implementation while preserving its core idea: extracting interprocedural vulnerability semantics via backward slicing. 
We use the tree-sitter library instead of Joern for static analysis and use its multi-language support to construct system dependence graphs for Python.
We run APPATCH once per vulnerability. 
When it outputs multiple candidate patches, we evaluate all candidates independently under the common patch application, build, and validation pipeline. 
APPATCH is considered to have produced a plausible patch for a vulnerability if at least one candidate meets the corresponding criterion.
In our experiments, APPATCH uses Claude Sonnet 4, the model \texttt{claude-sonnet-4-20250514}, as the backbone, with temperature 0 and no maximum token limit.

\subsubsection{Directly Prompted LLMs.}
\label{sec:llm-approaches}
We evaluate two widely used closed-source LLMs, GPT-4o
(\texttt{gpt-4o-2024-08-06}) and Claude Sonnet~4
(\texttt{claude-sonnet-4-20250514}), both decoded at temperature~0
to reduce sampling variance.

% and no output limit for deterministic results.

\textit{Prompt template.} We use a standardized prompt template, shown in Listing~\ref{lst:prompt-template}, that gives the model the vulnerable source file together with its repo-relative path and asks for a unified diff, forbidding edits to files it was not shown. 
The template also carries two hint blocks: a focus hint with the CVE identifier, and a localization hint that lists the non-test files and line ranges the human fix touched, explicitly marked as indicating where to focus but not how to fix. 

% \begin{lstlisting}[style=promptblock, language=Prompt, caption={Prompt template used for directly prompted LLMs.}, label={lst:prompt-template}]
% \begin{figure*}
% \centering
% \begin{minipage}{\textwidth}
\begin{lstlisting}[style=promptblock, language=Prompt, caption={Prompt template used for directly prompted LLMs.}, label={lst:prompt-template}]
# ROLE
You are a precise software engineer and security patcher.

# TASK
Given ONLY the source file(s) below (the only source of truth), produce a patch that removes the vulnerability
while preserving intended functionality.

# OUTPUT RULES (READ CAREFULLY)
- Output ONLY a unified diff patch (git-style preferred: 'diff --git ...'). No explanations. No markdown.
- The patch MUST apply cleanly relative to the repo-root using the provided repo-relative paths.
- Do NOT modify any files not provided.
- If no change is needed, output exactly:
  [No Patch]

# FOCUS HINTS
- ID: {cve_id}

# VULNERABILITY LOCALIZATION HINTS
The following non-test files and line ranges were changed in the human patch.
They indicate WHERE to focus, but not HOW to fix:
- {repo_relative_path} (short name: {file_name})
  - modify_or_remove lines: {modified_line_ranges}
  - add lines: {added_line_ranges}

# INPUT FILES (repo-relative path + full content)
{files_blob}

# NOW OUTPUT ONLY THE UNIFIED DIFF PATCH (or [No Patch]).
\end{lstlisting}
% \end{minipage}
% \end{figure*}

For any proposed patch, we apply it to the vulnerable project and run the paired oracle.
Patch plausibility is assessed via the paired oracle on \bench. We additionally assess patch correctness via manual inspection.

\subsubsection{Software Engineering Agents.}
We also evaluate three software engineering agents: OpenHands, Trae Agent, and SWE-agent.
These systems are not designed specifically for vulnerability repair, but they provide autonomous code editing and project-level reasoning that may allow them to fix vulnerabilities through iterative testing and refinement.

\textit{Agent frameworks.}
OpenHands offers a flexible environment where an agent can inspect files, run shell commands, and apply code edits.
Trae Agent enforces a lightweight test-driven repair loop, using build and test feedback to iteratively refine changes.
SWE-agent operates on a structured agent-computer interface that lets the model navigate and edit a real project tree.
We select these frameworks because they are widely used, end-to-end executable on real repositories, and they cover different agent designs.
All agents operate directly on the full project directory for each vulnerability instance.

\textit{Configuration.}
To ensure fairness, all agents use the same configuration:
i) Claude Sonnet 4 as the backbone model,
ii) temperature 0,
iii) no internet access,
iv) a maximum of 100 agent steps, and
v) a maximum API cost of 3 US dollars per vulnerability.
We use the default settings of each agent framework, modifying only the task instructions so that they follow the benchmark rules.
One exception applies to the cost cap: Trae Agent does not expose a spending limit per run, so for it we can only enforce the bound of 100 steps; as a consequence, its mean spend per instance exceeds the budget of 3 dollars (Table~\ref{tab:rq1-vul4py}), whereas the other agents stay within it.

\textit{Task setup.}
For each vulnerability instance, we provide the agent with 
i) the project workspace, ii) the same hints given to directly prompted LLMs, including the CVE identifier and the non-test files and line ranges changed by the human fix, iii) benchmark-specific build and validation commands, and iv) strict constraints intended to prevent leakage of ground truth fixes and to preserve the benchmark's intended evaluation pipeline.
Agents are instructed to produce a minimal patch that a) applies cleanly, b) passes the project build and setup step, and c) passes the benchmark's paired oracle, using only the provided build and test instructions.
Agents are explicitly forbidden from modifying or deleting tests, introducing alternative build systems, or using version control history such as \texttt{git log} or \texttt{git show}.
They may inspect and edit only repository source files and may run shell commands necessary for building and testing.
We evaluate only the final patch produced at termination, either success or budget exhaustion.
We do not consider intermediate patches generated during the agent trajectory.

\textit{Artifacts and collection.}
At the end of each run, we collect a unified diff of the final changes relative to the repository root, and the full contents of all modified non-test source files.
These artifacts are then passed to the same patch application, build, and validation pipeline used for directly prompted LLMs and APPATCH, ensuring consistent evaluation across approach categories.

\subsection{Evaluation Metrics}
We assess generated patches using the following metrics:

\begin{itemize}[left=0pt, labelsep=0.5em, itemsep=0pt]
\item \textbf{Security Test Passes.} Number of generated patches whose exploit (security) test passes on the patched revision, that is, what an exploit-only oracle would accept, without requiring the functional test suite to keep passing.
\item \textbf{Plausible Patches.} Number of generated patches that pass the benchmark's paired oracle, that is, the project functional test suite continues to pass and the exploit test passes on the patched vulnerable revision.
\item \textbf{Correct Patches.} Number of generated patches judged correct via manual inspection, compared to the corresponding human patches. 
Two authors independently inspect each plausible patch and label it correct only if it is semantically equivalent to the corresponding human fix. 
They reached an agreement of $\kappa = 0.67$ before discussion.
Disagreements are resolved by discussion.
\item \textbf{Cost.} Mean API spend per instance, in US dollars.
\item \textbf{Runtime.} Mean wall-clock seconds per instance.
\end{itemize}
Python's dynamic typing makes a separate compilation check uninformative, so we do not report a compilation rate.

\section{Experimental Results}
\label{sec:results}

% \subsection{RQ1: What are the characteristics of \bench as a dataset for automated Python vulnerability repair?}

\subsection{RQ1: Effectiveness of LLM-based approaches}
We evaluate the six approaches of Section~\ref{sec:setup} on \bench, measuring how many patches are plausible and how many are manually confirmed correct, alongside their cost and runtime.

\begin{table}[t]
\centering
\caption{Repair effectiveness, cost, and runtime on \bench. Sec.\ is the count passing the exploit (security) test alone, that is, what an exploit-only oracle would accept; Plau.\ is the paired oracle pass count, which additionally requires the project functional test suite to keep passing; Corr.\ is the subset judged to fix the underlying vulnerability by manual inspection against the upstream human fix. Cost and runtime are mean per-instance figures for patch generation.}
\label{tab:rq1-vul4py}
\small
\begin{tabular}{lrrrrr}
\toprule
\textbf{Approach} & \textbf{Avg.\ Cost} & \textbf{Avg.\ Time} & \textbf{Sec.} & \textbf{Plau.} & \textbf{Corr.} \\
\midrule
APPATCH      & \$0.05 &   93s &  3 &  2 &  2 \\
GPT-4o       & \$0.03 &    7s &  0 &  0 &  0 \\
Claude Sonnet 4 & \$0.05 & 6s &  5 &  5 &  4 \\
OpenHands    & \$1.32 &  490s & \textbf{51} & \textbf{43} & \textbf{41} \\
Trae Agent   & \$7.52 & 5,763s & 33 & 30 & 29 \\
SWE-agent    & \$1.43 & 2,052s & 27 & 24 & 22 \\
\bottomrule
\end{tabular}
\end{table}

Table~\ref{tab:rq1-vul4py} reports plausibility, manually verified correctness, mean cost, and mean runtime per instance for each approach over 100 \bench instances.
The ranking separates sharply by approach, and the same ordering holds for both plausibility and correctness.
The three software engineering agents dominate: OpenHands produces 43 out of 100 plausible patches, that is, 43\%, of which 41 are confirmed correct; Trae Agent 30 plausible and 29 correct; and SWE-agent 24 plausible and 22 correct.
The remaining approaches cluster near zero: Claude Sonnet 4 produces 5 plausible and 4 correct, APPATCH 2 plausible and 2 correct, and GPT-4o none.

\textit{Where the gap comes from.}
The ordering is striking because all six approaches build on the same class of backbone model and receive the same fix localization hints, yet every agent resolves at least five times as many vulnerabilities as the strongest non-agent approach, and OpenHands roughly ten times as many.
% yet the agents resolve an order of magnitude more vulnerabilities than direct prompting and the specialized tool. 
The separation tracks how much each approach can interact with the project rather than how strong its underlying model is. 
Direct prompting must emit a correct patch in a single shot from static context, and GPT-4o produces no surviving patch at all while Claude Sonnet 4 manages only 5 plausible; APPATCH adds slicing-based vulnerability context but still operates without execution feedback and yields 2 plausible patches. 
The agents, by contrast, can run the project, observe failing tests, and revise their edits before committing, which is precisely what the paired oracle rewards. 
This explains why OpenHands, the lowest-cost agent, nevertheless leads in coverage: iteration against concrete test signals matters more than per-step model effort. 
It also clarifies why the gains do not come for free, since the same iterative execution that drives coverage is what makes the agents far more expensive per instance.

\textit{Cost.}
Agents are roughly 26 to 150 times more expensive than direct prompting on a per-instance basis, and Trae Agent's iteration depth makes it both the slowest and the costliest. Because Trae Agent provides no hard cost cap and is bounded only by the 100-step limit, its mean per-instance spend exceeds the 3 dollar budget we set on the other agents, which partly explains its position as an outlier on cost.
Per plausible patch, however, the agents remain the only approaches that deliver coverage at all: direct prompting and the specialized tool are cheap and fast but yield few patches that survive the paired oracle.
The right operating point depends on whether absolute coverage or the raw cost per attempt matters most.

\begin{tcolorbox}[enhanced, size=fbox, drop fuzzy shadow southeast, boxrule=0.1pt]
\smallskip
\textbf{\textit{Answer to RQ1.}} Repair effectiveness splits sharply by approach. The three software engineering agents dominate, with OpenHands, Trae Agent, and SWE-agent producing 41, 29, and 22 correct patches, respectively, while the directly prompted LLMs and the specialized tool stay in the single digits: Claude Sonnet 4 with 4, APPATCH with 2, and GPT-4o with none, a five- to tenfold gap on a shared backbone.
% while the directly prompted LLMs and the specialized tool trail by an order of magnitude, Claude Sonnet 4 with 4, APPATCH with 2, and GPT-4o with none. 
The decisive factor is the agentic process of executing and iterating against project test signals, and this capability comes at a cost premium of roughly 26 to 150 times per instance.
\end{tcolorbox}

\subsection{RQ2: Reliability of the paired oracle}

\textit{Exploit test versus the full paired oracle.}
The Sec.\ column in Table~\ref{tab:rq1-vul4py} reports how many patches pass the exploit test on its own, before the functional test suite is taken into account.
Across all approaches, 119 patches clear the exploit test, but only 104 of them also leave the project functional test suite passing, so the functional half of the paired oracle removes 15 patches that pass the exploit test alone.
Of these 15, 14 fall on the agents, whose larger edits more often disturb unrelated behavior, and 1 on the specialized tool APPATCH: OpenHands drops from 51 to 43, Trae Agent from 33 to 30, SWE-agent from 27 to 24, and APPATCH from 3 to 2.

\textit{Plausibility is a high-precision proxy for correctness.}
A central concern with any oracle is whether passing it actually implies a correct fix. On \bench the paired oracle is strong on this axis: of the 104 plausible patches produced across all approaches, manual review
judged 98 to be semantically equivalent to the developer patch, a precision of
94\%. The remaining 6 were over-fitted to the tests: they pass the exploit
oracle but omit conditions that the developer patch handles and that the
tests do not exercise. The per-approach plausible to correct drop is correspondingly small: OpenHands 43 to 41, Trae Agent 30 to 29, SWE-agent 24 to 22, Claude Sonnet 4 5 to 4, and APPATCH 2 to 2. 

\textit{Reading the two error modes.}
The paired oracle can make mistakes in two directions, and both stay small on \bench. 
It can be too strict, rejecting a genuinely correct patch because the patch affects an unrelated functional test; the 15 patches removed by the functional half are dominated by the agents, whose broader edits occasionally touch behavior the functional suite happens to cover, so a fraction of that drop reflects test brittleness. 
It can also miss incorrect patches, admitting a patch that passes both tests yet does not truly fix the vulnerability; the 6 over-fitted patches fall here, satisfying the exploit and functional tests while omitting conditions the developer fix handles but the tests never exercise. 
The practical takeaway is that the exploit test alone is insufficient, since 12.6\% of exploit-passing patches fail the project's own functional suite, while pairing it with the functional test recovers a 94\% precision proxy whose residual error is concentrated in a handful of test-shaped shortcuts. 
This argues for the paired design as the default acceptance criterion for Python AVR and for treating manual confirmation as a complement rather than a replacement when the underlying vulnerability is subtle.

\begin{tcolorbox}[enhanced, size=fbox, drop fuzzy shadow southeast, boxrule=0.1pt]
\smallskip
\textbf{\textit{Answer to RQ2.}} The paired oracle is a reliable proxy for correct repair on two counts. 
First, the functional test is discriminating: of the 119 patches that pass the exploit test alone, only 104 also keep the project functional test suite passing, so the functional half removes 15 patches, 12.6\%, that an exploit-only oracle would have accepted without further checking, most of them from agents whose larger edits disturb unrelated behavior. 
Second, surviving the paired oracle is highly predictive of true correctness: 98 of the 104 plausible patches, 94\%, were manually confirmed semantically equivalent to the developer fix, with only 6 over-fitted to the tests. 
The paired oracle therefore meaningfully filters incorrect patches while rarely admitting oracle-targeted shortcuts.
\end{tcolorbox}

% \begin{tcolorbox}[enhanced, size=fbox, drop fuzzy shadow southeast, boxrule=0.1pt]
% \textbf{\textit{Summary of results.}} Under the paired oracle, repair effectiveness splits cleanly by approach: software engineering agents (OpenHands 41, Trae 29, SWE 22 correct) outperform directly prompted LLMs and the specialized tool (Claude 4, APPATCH 2, GPT-4o 0 correct) by an order of magnitude, despite sharing the same backbone model and receiving the same fix localization hints. The advantage is in the process: the ability to execute and iterate on project test signals, rather than model-driven. The paired oracle is also a high-precision proxy for correctness: 98 of 104 plausible patches were manually confirmed correct, so passing it rarely reflects an oracle-targeted shortcut.
% \end{tcolorbox}

\section{Threats to Validity}
\label{sec:threats}

\subsection{Construct Validity}
Our headline metric is the paired oracle, the project functional test suite together with the exploit test that the upstream fix commit ships. 
The main threat is that an oracle is unrelated to the vulnerability: the shipped test may exercise a benign new code path, depend on unreproducible external state, or be brittle to dependency drift, in which case every approach's success is silently inflated. 
We remove these cases during curation. 
Each accepted instance passes the manual oracle validation of Section~\ref{sec:construction}, where we confirm a clear correspondence between the added or modified test and the reported flaw and discard coincidental test edits.
% and our harness automatically flags non-discriminating functional tests, infrastructure errors, and dependency drift so they are reviewed rather than scored as repairs. 
Reproducibility is then a hard admission gate: an instance is admitted only if the exploit test fails on the vulnerable revision and passes on the fixed revision while the functional test passes on both, so instances whose oracle does not exhibit the expected before-and-after behavior are excluded.

\subsection{Internal Validity}

\paragraph{Adapted baseline.}
APPATCH was released for C and C++ and relies on Joern for static analysis.
Our Python port preserves its core idea, interprocedural vulnerability semantics via backward slicing, but substitutes tree-sitter for Joern. 
Its low score therefore measures our port under our prompt and budget, not APPATCH as published, and we make no claim about the original system's effectiveness on the languages it targets.
% We mitigate this threat by carefully adapting the implementation.

\paragraph{Leakage of fix commits into LLM training data.}
Every fix commit in \bench is public on GitHub and may have been seen during model pre-training, which could let a model recall a fix rather than reason to it. We cannot retract public data, but two observations bound the effect. 
First, the directly prompted LLMs fail by producing almost no patch that passes the paired oracle at all, 5 and 0 of 100 even when given line-level fix localization. Second, the agents outperform direct prompting while sharing the same backbone model, so the advantage tracks the ability to execute and iterate on test feedback rather than recall of a memorized fix. 

% \paragraph{Manual correctness labeling.}
% Correctness is a human judgment. 
% Two of the authors labeled each plausible patch independently, with agreement of $\kappa = 0.67$ before discussion, and all disagreements were resolved by discussion. 

\subsection{External Validity}
\bench covers 100 vulnerabilities across 60 open-source projects and 60 distinct CWEs spanning 2017 to 2025, but coverage of specialized Python such as asyncio internals, native C extensions, and machine learning serialization is uneven, and the reproducer filter favors web framework and serialization libraries. We mitigate over-generalization by scoping every claim to open-source Python with reproducible oracles and making no claim about closed-source or industrial codebases, and by releasing the construction pipeline so the benchmark can be extended toward underrepresented categories as new fixes containing tests become available.

% new
Another threat is the narrow set of systems we evaluate. 
We test one specialized repair tool, two directly prompted LLMs, and three software engineering agents, and every LLM-backed approach runs on a commercial, closed-weight model. Our results therefore describe what proprietary models can do today rather than what repair systems can do in general, and the process advantage we find for agents may not hold for open-weight models that follow instructions or use tools less reliably. 
We keep the backbone model fixed across the agent and direct prompting conditions so the comparison isolates process from model, and we release the harness with a documented interface so that open-weight models and other repair tools can be scored on the same oracle.

\section{Conclusion and Future Work}
\label{sec:conclusion}
We presented \textsc{Vul4Py}, a Python AVR benchmark of 100 real
vulnerabilities from 60 projects and 60 CWEs, spanning 2017 to 2025, each
carrying both an exploit and a functional oracle. Across six approaches, repair
effectiveness splits by paradigm rather than by model: 
on a shared backbone, the agents pass the paired oracle on 24 to 43 of 100 instances against at most 5 for direct prompting and the specialized tool, at 26 to 150 times the cost of prompting the same model directly.
% on a shared backbone, the agents pass an order of magnitude more paired oracles than direct prompting or the specialized tool, at 25 to 150 times the cost per instance. 
The functional oracle is not redundant either, rejecting one in eight patches that an exploit-only oracle would have accepted.

Three directions follow. First, our manual review covers only the 104 patches
that survive the paired oracle; labeling the roughly 500 failed attempts by
cause would turn pass rates into concrete diagnoses. Second, scaling past 100
instances will need either looser automation for finding fixes that ship tests,
or community-contributed oracles, for which we release the curation harness as a
quality gate. Third, every approach we evaluate runs on a closed-weight
commercial model, so whether the process advantage of agents holds on
open-weight backbones remains open.

\section*{Data Availability}
\bench is publicly available at \url{https://github.com/tabudz/vul4py}.

\bibliographystyle{splncs04}
\bibliography{main}

\end{document}